# Interactive evaluation of operation efficiency of the city's transport system by methods of *U*-statistics

O. D. Polishchuk, M. S. Yadzhak

Laboratory of Modeling and Optimization of Complex Systems
Pidstryhach Institute for Applied Problems of Mechanics and Mathematics, National Academy of Sciences of Ukraine,
Lviv, Ukraine
od_polishchuk@ukr.net

**Abstract** – Method of *U*-statistics is used to analyze the efficiency of functioning of the motor transport system of a large city as a complex network system with partially ordered traffic flows. Based on the results of continuous monitoring of movement of the ordered part of flows equipped with GPS-trackers, namely, public transport means, methods of interactive, forecasting and aggregated analysis of the state and operation process of motor transport system components of various hierarchy levels have been developed. The proposed technique can be easily automated and used for operational analysis and forecasting of the development of traffic situations on city highways and creation of effective tools to optimize the process of motor transport system functioning.

**Key words**: complex network, network system, road system, *U*-statistics, evaluation, forecasting, aggregation

**Вступ**. У працях [1–4] на прикладі залізничної транспортної системи була розроблена методика комплексного оцінювання стану та процесу функціонування мережевої системи (МС) із повністю впорядкованим рухом потоків. Такі системи складають значну частку сучасного індустріального суспільства. Однак, не менш важливими у функціонуванні економіки та забезпеченні життєдіяльності людського соціуму є системи із частково впорядкованим рухом потоків, тобто системи, у яких рух лише частини потоків є підпорядкованим певному графіку. Одним із показових прикладів таких систем є автотранспортні системи (АТС) великих міст. Задачі підвищення ефективності їх функціонування, оптимізації дорожньої інфраструктури та регулювання руху, а також вирішення проблеми заторів та запобігання дорожньо-транспортним пригодам і оперативне подолання їх наслідків давно привертають увагу дослідників [5–8]. Натепер для розв'язання цих задач все активніше залучаються можливості новітніх мобільних та інформаційних технологій [9–12]. Рух частини потоків у таких МС, а саме засобів громадського транспорту (ЗГТ), здійснюється згідно з наперед встановленим графіком. Очевидно, що оцінювання стану та ефективності функціонування таких систем є не менш важливим, ніж оцінювання МС, рух потоків у яких є повністю детермінованим. Натепер створено чимало математичних моделей процесу функціонування АТС великого міста: мікро- та макроскопічних, кінематичних, гравітаційних, ентропійних, які ґрунтуються на теорії станів, клітинних автоматів, конкурентних можливостей [13, 14] тощо. Загалом ці моделі достатньо адекватно відображають процеси, які перебігають в автотранспортних системах у некритичних умовах їх функціонування, але не можуть вирішити проблему критичного завантаження шляхів у години пік, визначити причини виникнення заторів та розв'язати інші важливі задачі моделювання та оптимізації автомобільного трафіку у місті.

**Мета статті** – розроблення інтерактивної методики оцінювання і прогнозування стану та ефективності функціонування елементів автотранспортної системи міста на основі методу *U*-статистик, який успішно застосовувався для верифікації достовірності прогнозів поведінки складних систем, адекватності їх математичних моделей [15, 16] тощо.

**Автотранспортна система міста, як мережева система із частково впорядкованим рухом потоків.** Дослідження АТС, як і будь-якої мережевої системи, починається із визначення її структури, у якості основних елементів якої виділяємо перехрестя (вузли) та поєднуючі їх ділянки шляху (ребра). Впорядкованою частиною потоків у цій системі вважатимемо засоби громадського транспорту. Сучасні методи GPS-моніторингу автотранспортних засобів дають можливість відстежувати у режимі реального часу (оновлення інформації про кожний об'єкт моніторингу відбувається кожні 15 сек.) місцезнаходження транспортного засобу, визначати його швидкість, місце та тривалість зупинок, перетин контрольних зон, найменші відхилення від заданого маршруту [10] тощо. На підставі цих даних системи GPS-моніторингу дозволяють формувати детальні звіти про пересування конкретного транспортного засобу, які можна використовувати як для оцінювання його стану, так і в сукупності для подальшого об'єктивного аналізу стану та якості функціонування автотранспортних мереж. Згадані технології дозволяють достатньо точно визначати щільність, інтенсивність та об'єми транспортних потоків, які пересуваються ребрами автотраспортної мережі (АТМ). Аналіз цих даних, одержаних для системи громадського транспорту, у якій зараз широко впроваджуються GPS-трекери, дозволяє робити у режимі реального часу опосередковані, але достатньо обґрунтовані висновки про стан та якість функціонування мережі. Більш того, процес одержання таких висновків та вироблення адекватних реакцій на транспортні ситуації можна легко автоматизувати.

Кожний власник ЗГТ зацікавлений в максимізації кількості перевезених пасажирів. Це відображається на області покриття маршрутами ЗГТ території міста та графіку руху кожного автобуса маршруту, який формується на основі дотримання, залежно від часу доби, близької до максимально допустимої швидкості його руху між зупинками та мінімізації часу посадки-висадки пасажирів на зупинках, тобто збільшенні кількості проходжень маршрутом кожного автобуса за добу. Невідповідним чином складений графік руху може призвести до скупчення автобусів на невеликій ділянці маршруту і унаслідок цього – до зменшення кількості пасажирів та доходів перевізника. Зазвичай маршрути громадського транспорту достатньо щільно охоплюють територію кожного великого міста, у тому числі усі основні його автошляхи.

Серед основних причин неефективного функціонування автотранспортної системи міста можна назвати вади дорожньої інфраструктури, наприклад, відсутність зручних розв'язок, незадовільний стан шляхів або транспортних засобів, низьку пропускну здатність доріг або надмірну щільність транспортних потоків, а також неефективну організацію їх руху за допомогою світлофорів. Застосування методів інтерактивного оцінювання поведінки систем із повністю впорядкованим рухом потоків, як засобу неперервного моніторингу функціонування АТС [4], виявляється не зовсім зручним із-за наявності великої кількості негативних випадкових впливів (ДТП, кліматичні умови і т. ін.), які порушують усталений процес функціонування автотранспортної системи міста.

**Методи *U*-статистик локального оцінювання стану та якості функціонування автотранспортних систем.** Для аналізу результатів неперервного моніторингу АТС можна використовувати низку статистичних методів: порівняння інтенсивності руху та швидкості у контрольних точках мережі, дисперсійний та кореляційний аналіз, визначення залежності між інтенсивністю, швидкістю та щільністю руху або між часом руху та зупинки на основі двохкомпонентних моделей кінематичної теорії транспортного потоку [12–14] і т. ін. Застосуємо для побудови інтерактивно-статистичного методу оцінювання процесу функціонування елементів АТС, як засобу аналізу результатів їх неперервного моніторингу, теорію *U*-статистик.

У якості об'єктів моніторингу оберемо регульовані та нерегульовані перехрестя автотранспортної мережі, а також елементарні ділянки шляху, які визначимо як складові ребер АТМ, обмежені перехрестями, світлофорами та зупинками ЗГТ. Дотримання встановленого графіка руху засобів громадського транспорту насамперед залежить від швидкості їх проходження цими елементами АТМ. Якщо середня швидкість руху автотранспорту елементарними ділянками у ненавантажені годині доби опосередковано, але від того не менш ґрунтовно, свідчить про стан цих ділянок (якість покриття), то перетин регульованих перехресть – про ефективність організації руху мережею (режиму роботи світлофорів).

При якісному автодорожньому покритті, враховуючи сучасні правила руху у населених пунктах, середня швидкість автотранспортних засобів, включаючи ЗГТ, на елементарних ділянках шляху повинна знаходитись у межах 40 – 50 км/год, принаймні у ненавантажені години доби. Ефективна організація роботи світлофорів (за виключенням заторів) означає затримку автотранспортного засобу на перехресті, яка не перевищує часу, протягом якого для напрямку його руху горить червоне світло. Це означає, що у якості очікуваних показників стану елементарних ділянок можна обрати середню швидкість руху на них ЗГТ, а у якості показників ефективності функціонування – значення режиму роботи світлофорів, які мінімізують час затримки на перехресті.

Локальне оцінювання процесу функціонування елементів АТС здійснюватимемо подобово, взявши за початок відліку 00:00 год. поточної доби. Позначимо через $v_{ij}^a$ – середню швидкість руху на $i$-тій елементарній ділянці $j$-го з початку доби ЗГТ, $1 \leq j \leq N_i$, де $N_i$ – максимальна кількість засобів громадського транспорту, які проходять цією ділянкою протягом доби. Нехай $v_{ij}^h$ – середня очікувана або обчислена на основі попередніх статистичних досліджень швидкість руху на $i$-тій ділянці автотранспортних засобів у момент $t_j \in [00:00, 24:00]$, яка визначається правилами дорожнього руху, просторовою геометрією (пряма, крива, одна або більше смуг) та довжиною ділянки, прогнозованою щільністю руху у даний час доби, кліматичними умовами (рис. 1) тощо. Для уніфікації процедур оцінювання руху ЗГТ на елементарних ділянках шляху та перехрестях АТМ визначимо сукупності $\mathbf{t}_i^{a,j} = \{t_{ik}^a\}_{k=1}^j$, $\mathbf{t}_i^{h,j} = \{t_{ik}^h\}_{k=1}^j$, у яких $t_{ik}^a = s_i/v_{ik}^a$, $t_{ik}^h = s_i/v_{ik}^h$ та $s_i$ – довжина $i$-тої елементарної ділянки, $i = \overline{1, L}$, де $L$ – кількість елементарних ділянок, які входять до складу АТМ міста та покриваються рухом громадського транспорту.

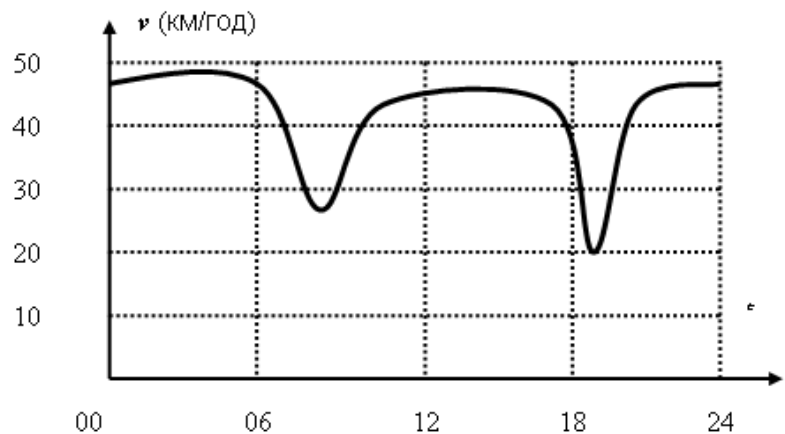

Рис. 1. Приклад очікуваної середньої швидкості руху автотранспортних засобів на $i$-тій елементарній ділянці АТС протягом доби

Тоді головна $U$-статистика $U_{ij}$ руху на $i$-тій ділянці АТМ, яка визначає відносне відхилення реального часу проходження ЗГТ ділянкою від очікуваного протягом періоду $[0, t_j]$, обчислюється за співвідношенням

$$U_{ij} = \left\| \mathbf{t}_i^{a,j} - \mathbf{t}_i^{h,j} \right\|_{R^j} \bigg/ \left( \left\| \mathbf{t}_i^{a,j} \right\|_{R^j} + \left\| \mathbf{t}_i^{h,j} \right\|_{R^j} \right), \qquad (1)$$

у якому $\|\mathbf{t}\|_{R^j}$ – норма евклідового простору $R^j$, породжувана скалярним добутком $<\mathbf{t},\boldsymbol{\tau}>_{R^j}=\sum_{k=1}^{j}t_k\tau_k$ .

Окрім головного значення, методи *U*-статистики містять також три додаткові критерії:

1) коефіцієнт частки зміщення $U_{ij}^m$, який оцінює систематичну похибку або наскільки середнє арифметичне значення реальних даних відрізняється від відповідного значення очікуваних показників на *i*-тій ділянці АТМ протягом періоду $[0,t_j]$ та обчислюється за співвідношенням

$$U_{ij}^m = j(\bar{t}_{ij}^a - \bar{t}_{ij}^h)^2 \Big/ \left\|\mathbf{t}_i^{a,j} - \mathbf{t}_i^{h,j}\right\|_{R^j}^2 , \qquad (2)$$

у якому $\bar{t}_{ij}^a = \sum_{k=1}^{j}t_{ik}^a$ та $\bar{t}_{ij}^h = \sum_{k=1}^{j}t_{ik}^h$ ;

2) коефіцієнт частки дисперсії $U_{ij}^s$, який оцінює міру співпадіння очікуваних та реальних швидкостей ЗГТ на *i*-тій ділянці АТМ протягом періоду $[0,t_j]$ та обчислюється за співвідношенням

$$U_{ij}^s = j(u_{ij}^a - u_{ij}^h)^2 \Big/ \left\|\mathbf{t}_i^{a,j} - \mathbf{t}_i^{h,j}\right\|_{R^j}^2 , \qquad (3)$$

у якому $u_{ij}^a = \left\|\mathbf{t}_i^{a,j} - \bar{\mathbf{t}}_i^{a,j}\right\|_{R^j} / j$, $\bar{\mathbf{t}}_i^{a,j} = \{\bar{t}_{ik}^a\}_{k=1}^j$, та $u_{ij}^h = \left\|\mathbf{t}_i^{h,j} - \bar{\mathbf{t}}_i^{h,j}\right\|_{R^j} / j$, $\bar{\mathbf{t}}_i^{h,j} = \{\bar{t}_{ik}^h\}_{k=1}^j$ ;

3) коефіцієнт частки коваріації $U_{ij}^c$, який оцінює залишкову похибку та дозволяє виділяти ті випадки, коли задовільні за першими двома критеріями реальні дані взаємно компенсують помилки спостережень на *i*-тій ділянці АТМ протягом періоду $[0,t_j]$ та обчислюється за співвідношенням

$$U_{ij}^c = 2j(1-r)u_{ij}^a u_{ij}^h \Big/ \left\|\mathbf{t}_i^{a,j} - \mathbf{t}_i^{h,j}\right\|_{R^j}^2 , \qquad (4)$$

де $r$ – коефіцієнт кореляції між реальними та очікуваними даними, який для систем такого типу зазвичай є близьким до 0 [14], $j=\overline{1,N_i}$, $i=\overline{1,L}$.

На основі обчислених значень визначених вище *U*-статистик сформуємо уточнені бальні оцінки їх поведінки протягом періоду $[0,t_j]$. Враховуючи, що значення цих статистик належать проміжку $[0, 1]$ та поведінка оцінюваних характеристик руху ЗГТ є тим кращою, чим ближчими є значення $U_{ij}$-, $U_{ij}^m$- та $U_{ij}^s$-статистик до 0, а значення $U_{ij}^c$-статистики – до 1, то інтерактивну статистичну уточнену бальну оцінку $e_{U_{ij}}(\mathbf{t}_i^{a,j})$ поведінки сукупності $\mathbf{t}_i^{a,j}=\{t_{ik}^a\}_{k=1}^j$ за головною $U_{ij}$-статистикою визначатимемо за співвідношенням

$$e_{U_{ij}}(\mathbf{t}_i^{a,j}) = \begin{cases} 5, & \text{якщо } U_{ij} \in [0.00, \gamma_1], \\ 4+4(\gamma_2 - U_{ij}), & \text{якщо } U_{ij} \in ]\gamma_1, \gamma_2], \\ 3+4(\gamma_3 - U_{ij}), & \text{якщо } U_{ij} \in ]\gamma_2, \gamma_3], \\ 2, & \text{якщо } U_{ij} \in ]\gamma_3, 1.00]. \end{cases} \qquad (5)$$

Аналогічно визначаються інтерактивні статистичні уточнені бальні оцінки $e_{U_{ij}^m}(\mathbf{t}_i^{a,j})$ та $e_{U_{ij}^s}(\mathbf{t}_i^{a,j})$ поведінки сукупності $\mathbf{t}_i^{a,j}$ за $U_{ij}^m$ та $U_{ij}^s$-статистиками відповідно, $j=\overline{1,N_i}$, $i=\overline{1,L}$.

Вважатимемо, що інтерактивна статистична оцінка $e_{U_{ij}^c}(\mathbf{t}_i^{a,j})$ поведінки сукупності $\mathbf{t}_i^{a,j}$ за $U_{ij}^c$-статистикою дорівнює:

$$e_{U_{ij}^c}(\mathbf{t}_i^{a,j}) = \begin{cases} 5, & \text{якщо } U_{ij}^c \in [\gamma_3, 1.00], \\ 4 + 4(U_{ij}^c - \gamma_2), & \text{якщо } U_{ij}^c \in [\gamma_2, \gamma_3[, \\ 3 + 4(U_{ij}^c - \gamma_1), & \text{якщо } U_{ij}^c \in [\gamma_1, \gamma_2[, \\ 2, & \text{якщо } U_{ij}^c \in [0.00, \gamma_1[, \end{cases} \quad j = \overline{1, N_i}, \ i = \overline{1, L}. \quad (6)$$

Значення інтерактивних статистичних оцінок, обчислені за формулами (5), (6), отримуються в уточненій бальній шкалі оцінювання [2], у якій основна понятійна оцінка, а саме «задовільно» та «добре», відповідної статистики уточнюється, виходячи з особливостей її поведінки протягом періоду часу $[0, t_j]$. Такі оцінки дозволяють скласти значно адекватніше уявлення про стан дорожнього полотна або ефективність організації руху автотранспортних потоків.

Поточний узагальнений висновок $E_{ij}(\mathbf{t}_i^{a,j})$ про поведінку сукупності $\mathbf{t}_i^{a,j}$ протягом періоду $[0, t_j]$ та відповідний фінальний узагальнений висновок $E_{i,N_i}(\mathbf{t}_i^{a,N_i})$ протягом доби за усіма $U$-статистиками отримуємо, використовуючи метод лінійної агрегації [3], а саме

$$E_{i,N_i}(\mathbf{t}_i^{a,N_i}) = (e_{U_{ij}}(\mathbf{t}_i^{a,N_i}) + e_{U_{ij}^m}(\mathbf{t}_i^{a,N_i}) + e_{U_{ij}^s}(\mathbf{t}_i^{a,N_i}) + e_{U_{ij}^c}(\mathbf{t}_i^{a,N_i}))/4, \ i = \overline{1, L}. \quad (7)$$

У таблиці 1 містяться результати оцінювання двох ділянок АТМ Львова. При цьому у (5) та (6) приймалося, що $\gamma_k = 0.25k, k = 1, 2, 3$. Ділянка 1 має чотири смуги і є горизонтальною прямою із асфальтовим покриттям та знаходиться у «спальному районі» на околиці міста, ділянка 2 має дві смуги і є горизонтальною прямою із покриттям бруківкою та знаходиться у середмісті Львова.

Для аналізу результатів неперервного моніторингу часу перетину ЗГТ регульованих перехресть, які знаходяться на шляху їхнього руху, методами $U$-статистик та обчислення відповідних уточнених бальних оцінок ефективності роботи світлофорів також застосовуються співвідношення (1)–(6), у яких у якості $\mathbf{t}_i^{a,j} = \{t_{ik}^a\}_{k=1}^j$ та $\mathbf{t}_i^{h,j} = \{t_{ik}^h\}_{k=1}^j$ приймаються реальні та очікувані значення часу затримки ЗГТ перед світлофором.

Таблиця 1

| Тип статистики / Узагальнена оцінка | Ділянка 1 | | Ділянка 2 | |
|---|---|---|---|---|
| | статистика | оцінка | статистика | оцінка |
| $U_{i,N_i}$ | 0.29 | 4.84 | 0.54 | 3.84 |
| $U_{i,N_i}^m$ | 0.21 | 4.52 | 0.51 | 3.96 |
| $U_{i,N_i}^s$ | 0.17 | 4.96 | 0.41 | 4.36 |
| $U_{i,N_i}^c$ | 0.62 | 3.48 | 0.08 | 2.00 |
| $E_{i,N_i}$, $i = 1, 2$ | 4.59 | | 3.54 | |

У таблиці 2 містяться результати оцінювання режиму роботи двох світлофорів, якими закінчувалися розглянуті у попередньому прикладі ділянки 1 та 2 відповідно. Для побудови фінального узагальненого висновку також використовувався метод лінійної агрегації.

Порівняння даних, які містяться у таблицях 1 та 2, дозволяє зробити висновок, що переваги, які надаються завдяки якісному стану шляху, можуть нівелюватися неефективним режимом роботи світлофорів та навпаки, що відкриває достатні резерви як удосконалення дорожньої інфраструктури, так і оптимізації процесу функціонування АТС міста з метою суттєвого покращення ефективності її роботи. Якщо метою оцінювання є виключно стан автошляхів, то визначені вище $U$-статистики доцільно обчислювати у ненавантажені періоди доби, наприклад, з 6:00 до 8:00, з 11:00 до 16:00 та з 21:00 до 24:00. Якщо ж метою оцінювання є виключно ефективність роботи світлофорів, то визначені вище $U$-статистики доцільно обчислювати у навантажені періоди доби, тобто з 8:00 до 10:00 та з 17:00 до 20:00. Відзначимо також, що для обчислення більш строгих оцінок $U$-статистик різних типів межі їх основних бальних (понятійних) оцінок можна корегувати, а для отримання більш адекватних узагальнених висновків – використовувати методи нелінійної або зваженої лінійної агрегації [3].

Таблиця 2

| Тип статистики / Узагальнена оцінка | Перехрестя 1 | | Перехрестя 2 | |
|---|---|---|---|---|
| | статистика | оцінка | статистика | оцінка |
| $U_{i,N_i}$ | 0.52 | 3.92 | 0.33 | 4.68 |
| $U_{i,N_i}^m$ | 0.37 | 4.52 | 0.27 | 4.92 |
| $U_{i,N_i}^s$ | 0.39 | 4.44 | 0.37 | 4.52 |
| $U_{i,N_i}^c$ | 0.24 | 2.00 | 0.36 | 3.44 |
| $E_{i,N_i}$, $i = 1, 2$ | 3.72 | | 4.39 | |

Проілюструємо корисність застосування визначених вище потокових $U$-статистик на наступному прикладі. У м. Львів, як і у більшості великих міст країни, існують сотні перехресть-вузлів зі структурним ступенем 4. Рух автотранспортних засобів на найбільш важливих із цих перехресть регулюється світлофорами. Неоптимальний режим роботи цих світлофорів, особливо в години пік, зазвичай стає причиною тривалих заторів. Значення $U$-статистик для вузлів АТС міста дозволяють насамперед виділити ті перехрестя, через які відбувається найбільш інтенсивний рух автотранспортних засобів та які потребують першочергової оптимізації роботи світлофорів. На рис. 2 зображена схема руху потоків на першому із згаданих вище перехресть (1 – «спальний» район міста до 150 тис. мешканців, 2 – замістя, 3 – промисловий район, 4 – середмістя).

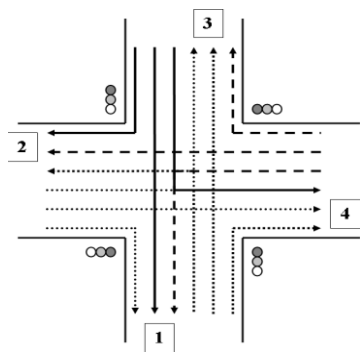

Рис. 2. Схема руху автотранспорту на перехресті великого міста

Очевидно, що у ранкові години пік (приблизно з 7:30 до 9:30) найбільші потоки спостерігаються зі «спального» району та замістя у напрямках промислової зони та середмістя, а у вечірні години пік (приблизно з 17:00 до 20:00) – у зворотних напрямках. Довший час «зелене» світло у кожному напрямку (протягом доби) тривало 15 сек., що призводило до постійних кількасотметрових заторів у ранкові та вечірні години пік. Це пояснювалось тим, що під час кожного «зеленого» світла через перехрестя пропускалося занадто мало транспортних засобів, а кількість під'їжджаючих до нього автомобілів поступово збільшувалась.

На основі кількісного аналізу накопичення автотранспортних засобів у різних напрямках у різні години доби під час очікування зеленого світла було встановлено найбільш оптимальний режим роботи світлофорів на цьому перехресті у робочі дні тижня (рис. 3). Лінія 1-2 відображає оптимальний режим роботи світлофорів для руху зі «спального» району та замістя, а лінія 3-4 – з боку промислового району та середмістя. Нажаль, забезпечити неперервну зміну режиму роботи світлофорів протягом доби не вдалося з технічних причин. Однак, було встановлено, що якщо тривалість «зеленого» світла у кожному напрямку дорівнює 25 сек., то це мінімізує накопичення об'ємів транспортних потоків, а, отже, протидіє виникненню заторів як у ранкові, так і у вечірні години пік.

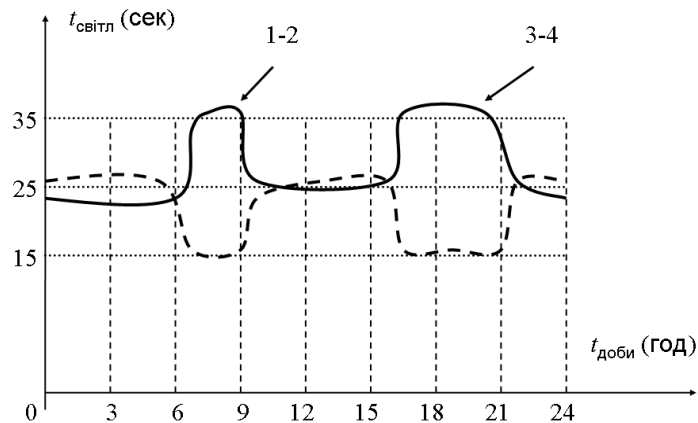

Рис. 3. Оптимальний режим роботи світлофорів на перехресті протягом доби

Подібні дослідження також були проведені для інших найбільш завантажених перехресть міста, що дозволило принаймні частково оптимізувати режими роботи світлофорів на них та суттєво зменшити проблему заторів за межами середмістя [17, 18].

Використовуючи апарат теорії $U$-статистик для дослідження руху автотранспорту нерегульованими перехрестями, можна визначити (у випадку постійних заторів) необхідність регулювання руху на таких перехрестях за допомогою світлофорів. Метою неперервного моніторингу АТС міста можуть бути і інші важливі показники її функціонування: місцезнаходження регулярних ДТП, ділянки постійних ремонтних робіт, невиправдані щільністю потоків затори, зменшення пропускної здатності ділянок шляху унаслідок масового паркування транспортних засобів і т. ін. У результаті таких досліджень світлофори встановлювались не лише на перехрестях, але й на пішохідних переходах на прямих ділянках дороги, на яких неодноразово здійснювалися наїзди.

**Агреговане оцінювання стану та якості функціонування автотранспортних систем.** Поділимо АТС міста на $R$ районів. Зазвичай ці райони є значно меншими за розміром, ніж однойменні адміністративні одиниці міста, та визначаються особливостями структури та процесу функціонування АТС. Позначимо через $G_r = (V_r, E_r)$ – структуру $r$-го району, у якій $V_r = \{v_{r,k}\}_{k=1}^{K_r}$ та $E_r = \{e_{r,k}\}_{k=1}^{N_r}$, $r = \overline{1, R}$, – сукупності вузлів та ребер цієї структури. Визначимо структуру кожного ребра $e_{r,k}$, як послідовність елементарних ділянок автошляху, розділених світлофорами та зупинками ЗГТ, тобто $e_{r,k} = \{e_{r,k}^l\}_{l=1}^{L_{r,k}}$. Побудову узагальнених висновків насамперед проводитимемо для сукупностей локальних оцінок, які характеризують стан ребер, підмереж районів та інфраструктури АТС міста загалом. Ці висновки формуються у результаті послідовного виконання наступних кроків.

1. Агрегація за часом (подобово) оцінок стану елементарної ділянки протягом певного періоду часу з використанням методу нелінійної агрегації, який у випадку рівноважливих оцінок елементів системи позбавлений недоліку методу лінійної агрегації, що полягає у нівелюванні як позитивних, так і негативних висновків стосовно їх стану та процесу функціонування [3]. Для спрощення викладу позначимо через $E_{l,m}^{r,k}$ – обчислений за формулою (7) узагальнений висновок про стан $l$-тої ділянки $k$-го ребра $r$-го району АТС, отриманий унаслідок неперервного моніторингу руху ЗГТ цією ділянкою під час $m$-тої доби, $m = \overline{1,M}$. Узагальнений висновок $E_{l,M}^{r,k}$ про стан досліджуваної ділянки протягом $M$ діб отримуємо методом нелінійної агрегації за співвідношенням

$$E_{l,M}^{r,k} = \prod_{m=1}^{M} E_{l,m}^{r,k} / (\varepsilon_{l,M}^{r,k})^{M-1}, \qquad (8)$$

де $\varepsilon_{l,M}^{r,k} = \sum_{m=1}^{M} E_{l,m}^{r,k} / M$.

Аналіз поведінки послідовності оцінок $E_{l,M}^{r,k}$ зі зростанням часового інтервалу, тобто значення $M$, дозволяє відстежувати швидкість погіршення стану ділянки та прогнозувати терміни її чергового ремонту [2].

2. Побудова методом нелінійної агрегації узагальненого висновку $E_M^{r,k}$ про стан ребра АТМ, яке складається з оцінених на попередньому кроці елементарних ділянок, а саме

$$E_M^{r,k} = \prod_{l=1}^{L_{r,k}} E_{l,M}^{r,k} / (\varepsilon_M^{r,k})^{L_k - 1}, \qquad (9)$$

де $\varepsilon_M^{r,k} = \sum_{l=1}^{L_{r,k}} E_{l,M}^{r,k} / L_k$.

3. Побудова методом гібридної агрегації [3], який поєднує переваги методів нелінійної та зваженої лінійної агрегації, узагальненого висновку $E_M^r$ про стан транспортної інфраструктури окремого району АТС міста з урахуванням пріоритетності окремих ребер АТМ. Поділимо усі ребра $r$-го району на $N_r$ груп ребер однакової пріоритетності $\rho_n^r$, $n = \overline{1,N_r}$, значення якої є пропорційними до об'ємів автотранспортних потоків, які проходять ребром за $M$ діб. Тоді узагальнений висновок $E_M^r$ про стан автошляхів району визначається за співвідношенням

$$E_M^r = \sum_{n=1}^{N_r} (\rho_n^r \widetilde{E}_M^{r,n}) / \sum_{n=1}^{N_r} \rho_n^r, \qquad (10)$$

у якому $\widetilde{E}_M^{r,n}$ – отриманий методом нелінійної агрегації узагальнений висновок для сукупності ребер $n$-тої групи, $n = \overline{1,N_r}$.

4. Побудова методом гібридної агрегації узагальненого висновку $E_M$ про стан транспортної інфраструктури АТС міста загалом з урахуванням пріоритетності підмереж окремих районів АТМ. Поділимо усі райони АТС міста на $N$ груп районів однакової пріоритетності $\sigma_n$, $n = \overline{1,N}$, значення якої є пропорційними до об'ємів автотранспортних потоків, які проходять районом за $M$ діб. Тоді узагальнений висновок $E_M$ про стан автошляхів міста визначається за співвідношенням

$$E_M = \sum_{n=1}^{N} (\sigma_n \widetilde{E}_M^n) / \sum_{n=1}^{N} \sigma_n, \qquad (11)$$

у якому $\widetilde{E}_M^n$ – отриманий методом нелінійної агрегації узагальнений висновок для сукупності районів *n*-тої групи, $n = \overline{1, N}$.

Показником достовірності отриманих агрегованих оцінок у межах міста є питома вага сукупності маршрутів громадського транспорту у структурі усіх його вулиць [19, 20]. Прогнозування поведінки оцінок (9)–(11) доцільно використовувати під час планування термінів та обсягів необхідних ремонтних робіт автошляхів міста, а також витрат, які їх супроводжують [3].

Наступний етап полягає у побудові узагальнених висновків для сукупностей локальних оцінок, які характеризують ефективність режимів роботи світлофорів на автошляхах міста. Ці висновки, аналогічно до оцінок стану, формуються у результаті послідовного виконання наступних кроків:

1) нелінійна агрегація за часом (подобово) оцінок режиму роботи окремого світлофора протягом певного періоду часу;

2) побудова методом нелінійної агрегації та прогноз поведінки узагальненого висновку про ефективність роботи усіх оцінених на попередньому кроці світлофорів окремого регульованого перехрестя АТМ;

3) побудова методом гібридної агрегації та прогноз поведінки узагальненого висновку про ефективність роботи усіх світлофорів району міста з урахуванням пріоритетності окремих його регульованих перехресть;

4) побудова методом гібридної агрегації та прогноз поведінки узагальненого висновку про ефективність роботи усіх світлофорів міста з урахуванням пріоритетності окремих його районів.

Показником достовірності отриманих агрегованих оцінок є питома вага транспортних потоків, які проходять через оцінену вище сукупність світлофорів, порівняно з усіма транспортними потоками міста. Сформовані узагальнені висновки створюють достатньо адекватне уявлення про ефективність організації руху транспортних засобів у місті та можуть бути використані для оптимізації режиму роботи світлофорів на його автошляхах [21].

Агреговане оцінювання також доцільно здійснювати з метою аналізу та прогнозування розвитку поточної автотранспортної ситуації, яка склалася на окремому маршруті руху ЗГТ, в районі міста або його АТС загалом. Узагальнені висновки цього типу формуються у результаті виконання наступних кроків:

1) побудова методом гібридної агрегації та прогноз поведінки узагальненого висновку про поточний стан (завантаженість) маршруту руху ЗГТ з урахуванням пріоритетності окремих ребер АТМ, які знаходяться на шляху цього маршруту;

2) побудова методом гібридної агрегації та прогноз поведінки узагальненого висновку про ефективність роботи світлофорів на маршруті руху ЗГТ у поточний момент часу з урахуванням пріоритетності окремих світлофорів АТМ, які знаходяться на шляху цього маршруту.

Агреговані оцінки можна будувати і для послідовності елементарних ділянок шляху та світлофорів, які утворюють маршрут руху ЗГТ:

1) побудова методом гібридної агрегації та прогноз поведінки узагальненого висновку про поточний стан (завантаженість) автошляхів окремого району міста з урахуванням пріоритетності ребер, які входять до складу АТМ району;

2) побудова методом гібридної агрегації та прогноз поведінки узагальненого висновку про ефективність роботи світлофорів окремого району міста з урахуванням пріоритетності світлофорів, які входять до складу АТМ району.

Агреговані оцінки також можна будувати для сукупності елементарних ділянок шляху та світлофорів, які входять до складу АТМ району:

1) побудова методом гібридної агрегації та прогноз поведінки узагальненого висновку про поточний стан (завантаженість) автошляхів міста загалом з урахуванням пріоритетності ребер, які входять до складу його АТМ;

2) побудова методом гібридної агрегації та прогноз поведінки узагальненого висновку про ефективність роботи світлофорів міста загалом з урахуванням пріоритетності світлофорів які входять до складу його АТМ.

Агреговані оцінки можна будувати і для сукупності елементарних ділянок шляху та світлофорів, які входять до складу АТМ міста [19-21].

**Висновки.** Розроблені у статті методи оцінювання та отримані результати можна успішно застосовувати для оцінювання поточних автотранспортних ситуацій на окремих ділянках міської автомережі та прогнозування у режимі реального часу їх подальшого розвитку. Для глибшого аналізу стану та ефективності функціонування АТС міста застосовано відповідні методи побудови узагальнених висновків, які дозволяють визначати найбільш завантажені у поточний момент або у найближчому майбутньому складові АТМ та завчасно обирати альтернативні шляхи руху в об'їзд цих складових, перерозподіляючи тим самим щільність транспортних потоків. Як і для систем із повністю впорядкованим рухом потоків, запропонована у цій статті методика також має комплексний характер, оскільки поєднує у собі взаємопов'язані методи інтерактивного, прогностичного та агрегованого оцінювання поведінки складових системи різного типу та рівнів ієрархії. Порівняно з іншими, ця методика достатньо просто реалізується за допомогою вже впроваджених у АТС більшості великих міст засобів GPS-трекінгу руху ЗГТ та легко автоматизується з використанням сучасних інформаційних технологій. Пропоновані методи потребують попереднього опрацювання в режимі реального часу значних обсягів вхідних даних, що можна здійснити, застосовуючи високопаралельні алгоритми цифрової фільтрації [22]. Окрім цього, з метою проведення досліджень АТС на сучасних обчислювальних засобах [23], необхідно на основі методів $U$-статистик будувати паралельні алгоритми оцінювання. Саме в цьому автори вбачають напрямок своїх подальших досліджень.

**Перелік посилань**

1. Поліщук Д. О. Комплексне детерміноване оцінювання складних ієрархічно-мережевих систем: І. Опис методики / Д. О. Поліщук, О. Д. Поліщук, М. С. Яджак // Системні дослідження та інформаційні технології. – 2015. – № 1. – С. 21-31.
2. Поліщук Д. О. Комплексне детерміноване оцінювання складних ієрархічно-мережевих систем: ІІ. Локальне та прогностичне оцінювання / Д. О. Поліщук, О. Д. Поліщук, М. С. Яджак // Системні дослідження та інформаційні технології. – 2015. – № 2. – С. 26-38.
3. Поліщук Д. О. Комплексне детерміноване оцінювання складних ієрархічно-мережевих систем: ІІІ. Агреговане оцінювання / Д. О. Поліщук, О. Д. Поліщук, М. С. Яджак // Системні дослідження та інформаційні технології. – 2015. – № 4. – С. 20-31.
4. Поліщук Д. О. Комплексне детерміноване оцінювання складних ієрархічно-мережевих систем: IV. Інтерактивне оцінювання / Д. О. Поліщук, О. Д. Поліщук, М. С. Яджак // Системні дослідження та інформаційні технології. – 2016. – № 1. – С. 7-16. https://doi.org/10.20535/SRIT.2308-8893.2016.1.01.
5. Kamenchukov A. Evaluation of road repair efficiency in terms of ensuring traffic quality and safety / A. Kamenchukov, V. Yarmolinsky, I. Pugachev // Transportation Research Procedia. – 2018. – Vol. 36. – P. 627-633. https://doi.org/10.1016/j.trpro.2018.12.142.
6. Su F. On urban road traffic state evaluation index system and method / F. Su, H. Dong, L. Jia, X. Sun // Modern Physics Letters B. – 2017. – Vol. 31. – No. 1. – 1650428. https://doi.org/10.1142/S0217984916504285.
7. Iwanowicz D. Analysis of the Methods of Traffic Evaluation at the Approaches of Urban Signalised Intersections / D. Iwanowicz, J. Chmielewski // In: Macioszek E., Kang N., Sierpiński G. (eds) Nodes in Transport Networks – Research, Data Analysis and Modelling. – Lecture Notes in Intelligent

8. Xiaoliang S. Research on Traffic State Evaluation Method for Urban Road / S. Xiaoliang, J. Jinke, Z. Jinjin, L. Jun // International Conference on Intelligent Transportation, Big Data and Smart City. – 2015. – P. 687-691. https://doi.org/10.1109/ICITBS.2015.174.
9. Lewandowski M. Road Traffic Monitoring System Based on Mobile Devices and Bluetooth Low Energy Beacons / M. Lewandowski, B. Płaczek, M. Bernas, P. Szymała // Wireless Communications and Mobile Computing.– 2018.– Vol. 2018.– 3251598.– 12 p. https://doi.org/10.1155/2018/3251598.
10. Внуков А. Б. Современные системы навигации и слежения за наземными транспортными средствами на базе спутниковых технологий / А. Б. Внуков // Горная промышленность. – 2006. – № 6. – С. 97-101.
11. Jin J. An intelligent control system for traffic lights with simulation-based evaluation / J. Jin, X. Ma, I. Kosonen // Control Engineering Practice. – 2017.– Vol. 58.– P. 24-33. https://doi.org/10.1016/j.conengprac.2016.09.009.
12. Кочерга В. Г. Интеллектуальные транспортные системы в дорожном движении / В. Г. Кочерга, В. В. Зырянов, В. И. Коноплянко. – Изд. Ростовского гос. строит. ун-та, Ростов на Дону, 2001. – 108 с.
13. Швецов В. И. Математическое моделирование транспортных потоков / В. И. Швецов // Автоматика и телемеханика. – 2003. – № 11. – С. 3-46.
14. Буслаев А. П. Вероятностные и имитационные подходы к оптимизации автодорожного движения / А. П. Буслаев, А. В. Новиков, В. М. Приходько, А. Г. Таташев, М. В. Яшина. – Мир, Москва, 2003. – 368 с.
15. Korolyuk V. S. Theory of U-Statistics / V. S. Korolyuk, Y. V. Borovskich. – Springer Science & Business Media, Berlin, 2013. – 554 p. https://doi.org/10.1007/978-94-017-3515-5.
16. Lee A. J. U-Statistics: Theory and Practice /A. J. Lee.– Routledge, London, 2019. – 320 p.
17. Polishchuk O. Issues of regional development and evaluation problems / O. Polishchuk, D. Polishchuk, M. Tyutyunnyk, M. Yadzhak // AASCIT Communications. – 2015. – Vol. 2. – No. 3. – P. 115-120.
18. Polishchuk A. D. About convergence the methods of projections for solution potential theory integral equation / A/ D/ Polishchuk // Computer centre of Siberian Division of AS of the USSR. – 1988. – Preprint 776. – 11 p.
19. Поліщук О. Д. Мережеві структури та системи: І. Потокові характеристики складних мереж / О. Д. Поліщук, М. С. Яджак // Системні дослідження та інформаційні техноло-гії. – 2018. – № 2. – С. 42-54.
20. Поліщук О. Д. Мережеві структури та системи: ІІ. Серцевини мереж та мультиплексів / О. Д. Поліщук, М. С. Яджак // Системні дослідження та інформаційні технології. – 2018. – № 3. – С. 38-51.
21. Поліщук О. Д. Мережеві структури та системи: ІІІ. Ієрархії та мережі / О. Д. Поліщук, М. С. Яджак // Системні дослідження та інформаційні технології. – 2018. – № 4. – С. 82-95.
22. Yadzhak M. S. On optimal in one class algorithm for solving three-dimensional digital filtering problem / M. S. Yadzhak // Journal of Automation and Information Sciences. – 2001. – Vol. 33. – No. 1. – P. 51-63.
23. Яджак М. С. Високопаралельні алгоритми та засоби для розв'язання задач масових арифметичних і логічних обчислень / М. С. Яджак // Київський національний університет ім. Тараса Шевченка.– Автореферат дисертації ... д. ф.-м. н., спеціальність 01.05.03.– 2009.– 33 с.